\documentclass[12pt,reqno]{article}
\usepackage{latexsym}
\usepackage{amsmath}
\usepackage[final]{pdfpages} 
\usepackage{graphicx}
\usepackage{amssymb}
\usepackage{euscript}
\usepackage{eufrak}
\usepackage{cancel}
\usepackage{fancyhdr}
\makeatletter\newlength{\earraycolsep}
\newlength{\textlarg}

\renewcommand{\thefootnote}{\alph{footnote}}
\begin{document}
\pagestyle{myheadings}

\title{Regge symmetry and partition of  Wigner $3$-$j$  \\  or super $3$-$j^S$ symbols: unknown properties \\}
\author{\tt Lionel Br\'{e}hamet \thanks {email: brehamet.l@orange.fr} \\ 
Research scientist \hspace{0.38em} 
\\ {France \hspace{4.92em} }
\date{}
}
\maketitle

\begin{abstract}
\noindent 
For each generic $(3\mbox{-}j)$ the column parities,  $2(j \pm m)$, define $3$ intrinsic parities: $\alpha, \beta,\gamma$.
In algebra $so(3)$ only $(3\mbox{-}j)_{\alpha}$ exists whereas super-algebra $osp(1|2)$ admits $3$ kinds of super-symbols 
$(3\mbox{-}j)^{S}_{\alpha}, (3\mbox{-}j)^{S}_{\beta}, (3\mbox{-}j)^{S}_{\gamma}$.
Instead of 4 for $\{6\mbox{-}j\}$ symbols, Regge symmetry this time produces 5 partitions
${\tt S}_{\boldsymbol{\wr}}(0), 
{\tt S}_{\boldsymbol{\wr}}(1), {\tt S}_{\boldsymbol{\wr}}(2), 
{\tt S}_{\boldsymbol{\wr}}(4), {\tt S}_{\boldsymbol{\wr}}(5)$, 
with ${\tt S}_{\boldsymbol{\wr}}(3) = \emptyset$. Valid for $(3\mbox{-}j)_{\alpha},
(3\mbox{-}j)^{S}_{\alpha,\gamma}$ they reduce to 2 for $(3\mbox{-}j)^{S}_{\beta}$ with ${\tt S}_{\boldsymbol{\wr}}(0), 
{\tt S}_{\boldsymbol{\wr}}(1)$. Unexpectedly a symbol $(3\mbox{-}j)^{S}_{\beta}$ and its 'Regge-transformed' may be opposite in sign.
In terms of integer parts and super-triangle $\triangle^{S}$ a formula similar to that of a $(3\mbox{-}j)$  is obtained for the $(3\mbox{-}j)^{S}$. Some forbidden $(3\mbox{-}j)^{S}_{\beta}$ require an analytic prolongation, consistent with Regge $\beta$-partitions.
\\ \\
\noindent 
PACS: 03.65.Fd  Algebraic methods\\  
PACS: 02.20-a  Group theory\\
PACS: 11.30.Pb  Supersymmetry  \\ [1em]
{\sl Keywords: Angular momentum in Quantum Mechanics, Wigner 3j-symbol, partition, Regge symmetry, 
Supersymmetry, Racah-Wigner calculus, $S3$-$j$, super $3$-$j^S$  symbol.}
\end{abstract}

\renewcommand{\thefootnote}{\arabic{footnote}}
\setcounter{footnote} {0}

\begin{flushleft}
\newpage
			\section{Introduction}\label{Intro}
As in our recent work on $\{6$-$j\}$ symbols partitions \cite{BrehametPJMPA} the aim is to carry out a similar analysis on 
$(3$-$j)$ symbols. The first task is  to find the right partition parameters for the $(3$-$j)$ symbols. 
{\it A priori}, they are far to be apparent data. However it well seems that they are involved in the analytic formulas 
themselves \cite{BrehametPJMPA}, here under the form of $(j_k\pm m_k)$ where $k$ refers to the kth column, $k=[1,3]$.
This key-parameter allows one to define a `column-parity' {\tt even} or {\tt odd}
according  to the parity of ${\tt 2}(j_k\pm m_k)$, respectively. Any column $\scriptsize{\tt c}_k=
\left| \begin{array}{c} j_k \\ m_k \end{array} \right|$ 
can be of two kinds, denoted by a shorthand notation like
$ \stackrel{|ev|}{{\tt c}_k}$ or $ \stackrel{|od|}{{\tt c}_k}$. 
In spite of its binary appearance, this is a  parameter different from the binary variable $2(j-l)$ introduced in 
\cite[p. 2477]{Daumensetal1} in relation to the $so(3)$ doublets $l=j, l=j-\frac{1}{2}$.
Thus alternative notations are possible:
\begin{equation}
\left(\begin{array}{ccc}j_1 & j_2 & j_3  \\
m_1 & m_2 & m_3 \end{array} \right)=
\left(\begin{array}{ccc}
\left| \begin{array}{c} j_1 \\ m_1 \end{array} \right|
&\left| \begin{array}{c} j_2 \\ m_2 \end{array} \right|
&\left| \begin{array}{c} j_3 \\ m_3 \end{array} \right|
\end{array} \right)
=({\tt c}_1\; {\tt c}_2\; {\tt c}_3).
\end{equation}
For $so(3)$ any $(3$-$j)$ is of kind $(\stackrel{|ev|}{{\tt c}_1}\; \stackrel{|ev|}{{\tt c}_2}\; \stackrel{|ev|}{{\tt c}_3})$.
This will be different for $osp(1|2)$ and $(3$-$j)^{S}$ symbols.\\
As will be seen further the concept of `column-parity' naturally leads to properly assign  intrinsic parities  to super $(3$-$j)^{S}$ symbols \cite{Daumensetal1,L.B.Nuov1.2006} and classify  their Regge-partitions. 
\\The paper is organized as follows:
sections \bf{\ref{StandardFormula}}-\bf{\ref{ReggeStandard}} \rm 
are devoted to $(3$-$j)$ symbols, \bf{\ref{SuperFormula}}-\bf{\ref{ReggeSuper}} \rm to super $(3\mbox{-}j)^{S}$  symbols
and \bf{\ref{AnalyticProlong}} \rm to an analytic prolongation of some `forbidden' super symbols.

				\section{Analytic formula for  $(3$-$j)$ symbols}\label{StandardFormula}
\renewcommand{\theequation}{\ref{StandardFormula}.\arabic{equation}}
For Wigner $3$-$j$ symbols, denoted here by $(3$-$j)$,  the most commonly used expression  \cite{Rot,Bied.Louck1,Edmonds,Iachello} can be written down as
\setcounter{equation}{0}
\begin{equation}\label {eq:Wigner3j}
\left(\begin{array}{ccc}j_1 & j_2 & j_3  \\
m_1 & m_2 & m_3 \end{array} \right)_{\!\!|\sum_k \! m_k\!=0}\!= \bigtriangleup(j_1 j_2 j_3 )\;
v\left(\begin{array}{ccc}j_1 & j_2 & j_3  \\
m_1 & m_2 & m_3 \end{array} \right)_{\!\!|\sum_k \! m_k\!=0},
\end{equation}
where $\bigtriangleup$ triangle of Edmonds  \cite[p. 99] {Edmonds} has been used here for convenience
\begin{equation}   \label{eq:TriangEdmup}
\bigtriangleup(a b c)= \begin{array} {c}\displaystyle
\left( \frac{(a+b-c)!(a-b+c)!(-a+b+c)!}
{(a+b+c+1)!} \right)^{1/2}\end{array},
\end{equation}
and $v$ is directly arranged with $(j_k\pm m_k)$ parameters announced in introduction:
\begin{eqnarray}\label {eq:v3jformulalb}
\lefteqn{
v\left(\begin{array}{ccc}j_1 & j_2 & j_3  \\
m_1 & m_2 & m_3 \end{array} \right)\!= \begin{array} {c}(-1)^{j_1+m_1-(j_2-m_2)}\end{array}
\scriptsize \begin{array} {c}\left( \prod_{k}(j_k+m_k)! (j_k-m_k)!  \right) ^{\frac{1}{2}}\end{array}\normalsize  
}\nonumber \\
&&  \times  
\begin{array}{c}{\displaystyle \sum_{z}} \frac{(-1)^{z}}{z!
(z -(j_2+m_2-(j_3-m_3)))!
(z-((j_1-m_1)-(j_3+m_3)))!
(j_1+ m_1+ j_2+ m_2 - (j_3- m_3))-z)!(j_1-m_1-z)!(j_2+m_2-z)!}\, \nonumber .
 \end{array}\\
\end{eqnarray}
This is nothing more than that used in Ref.  \cite{Rot}\footnote
{Misprints: in (1.11) no frontal phase, in rhs of (1.12), $m_3$ to be replaced by $m_2$.} for computing  $(3$-$j)$ symbols numerical values.

				\section{Regge symmetry of ($3$-$j$) symbols}\label{ReggeStandard}
				\renewcommand{\theequation}{\ref{ReggeStandard}.\arabic{equation}}
				\setcounter{equation}{0}
By ignoring  the phases, the symmetry group contains $12$ elements and temporarily will be denoted by  $S_{\boldsymbol{\wr}}$  (lack of better).
Thanks to Regge \cite{Regge} surprising symmetries became known since 1958 and relate to Wigner ($3$-$j$) symbols. 
They were already reported in Table of Rotenberg and al. \cite{Rot}, and analyzed in standard books  like  \cite{Bied.Louck1} where
by way of conclusion we only learn that the initial symmetry group $S_{\boldsymbol{\wr}}$  becomes a larger group of order $72$.\\
\hspace*{1em}According to our analysis done with $\{6$-$j\}$ symbols \cite{BrehametPJMPA}, we are interested in the production
of new triangles $(j'_1\, j'_2\, j'_3)$ from a given $(j_1\, j_2\, j_3)$. 
We shall write out in detail only the relevant transformations
by avoiding phase factors in formulas, which is possible using one of the twelve ($3$-$j$) symmetries. Our notations of the partition parameters 
will be the following
\begin{equation}\label{eq:PartitionParam}
\boldmath
j_k^{+}= (j_k\!+\!m_k), \quad j_k ^{-}= (j_k\!-\!m_k).
\end{equation}
As a matter of fact, a glance at the Regge array \cite{Rot,Bied.Louck1}, also used to represent a ($3$-$j$) symbol, directly shows the underlying existence of these parameters:
\begin{equation}\label{eq:ReggeArray}
\left(\begin{array}{ccc} j_1 & j_2 & j_3  \\ m_1 & m_2 & m_3 \end{array} \right)= R =
\left[ \begin{array}{ccc} -j_1\!+\!j_2\!+\!j_3 & \,j_1\!-\!j_2\!+\!j_3 &\, j_1\!+\!j_2\!-\!j_3  \\ j_1\!-\!m_1 & \,j_2\!-\!m_2 &\, j_3\!-\!m_3 \\
 \, j_1\!+\!m_1 & \,j_2\!+\!m_2 & \,j_3\!+\!m_3\end{array} \right],
\end{equation}
\begin{equation}\label{eq:ReggeArraybis}
R=
\left[\begin{array}{ccc} R_1^1 & R_1^2 & R_1^3  \\ R_2^1 & R_2^2 & R_2^3 
\\ R_3^1 & R_3^2 & R_3^3\end{array} \right] =
\left[ \begin{array}{ccc} - j_1^{-}\!+j_2 ^{+} \!+j_3 ^{+} & \quad j_1 ^{+}\!-j_2 ^{-}\!+j_3 ^{+} 
& \quad j_1 ^{+}\!+j_2 ^{+}\!-j_3^{-}  \\
 j_1 ^{-} & \,j_2 ^{-}&\,j_3 ^{-} \\
 j_1 ^{+} & \,j_2 ^{+} & \,j_3 ^{+} 
\end{array} \right].
\end{equation}

\hspace*{1em}Below $five$  Regge transformations are listed from  ${\cal{R}}_1$ up to ${\cal{R}}_5$.  They generate at most $five$ distinct triangles different from the original.  
This  means also $five \, distinct$  ($3$-$j$) symbols, of course {\tt with the same numerical value}. We emphasize this point because
for super $(3$-$j)^{S}$ symbols it may happen that numerical values do not have the same sign.\\ \vspace{0.5em}

			\hspace*{12em}{\tt Overview of Regge transformations}
\begin{eqnarray}
\left(\begin{array}{ccc} j_1 & j_2 & j_3  \\ m_1 & m_2 & m_3 \end{array} \right)
&=\left(\begin{array}{ccc}j_1 &\, \frac{1}{2}(\,j_3^{-}+\,j_2^{-} )& \, \frac{1}{2}(\,j_3^{+}\!+\,j_2^{+} )  \\ [0.25em]
( j_2\!-\!j_3)&\, \frac{1}{2}(\,j_3^{-}-\,j_2^{-} )& \, \frac{1}{2}(\,j_3^{+}\!-\,j_2^{+} )\end{array} \right), &
\qquad \qquad \qquad \qquad \qquad \;\; {\cal{R}}_1 \label{Regge1}\\ [0.25em]
\phantom{\left(\begin{array}{ccc} j_1 & j_2 & j_3  \\ m_1 & m_2 & m_3 \end{array} \right)}
&=\left(\begin{array}{ccc}\frac{1}{2}(\,j_1^{-}\!+j_3^{-} )&\, j_2 & \, \frac{1}{2}(\,j_1^{+}\!+j_3^{+} )  \\ [0.25em]
\frac{1}{2}(\,j_1^{-}\!-j_3^{-} )&\, ( j_3\!-\!j_1)& \, \frac{1}{2}(\,j_1^{+}\!-j_3^{+} )\end{array} \right), & 
\qquad\qquad\qquad\qquad\qquad\;\,{\cal{R}}_2 \label{Regge2}\\ [0.25em]
\phantom{\left(\begin{array}{ccc} j_1 & j_2 & j_3  \\ m_1 & m_2 & m_3 \end{array} \right)}
&=\left(\begin{array}{ccc}\frac{1}{2}(\,j_2^{-}\!+j_1^{-} ) &\, \frac{1}{2}(\,j_2^{+}\!+j_1^{+} )& \,   j_3   \\ [0.25em]
\frac{1}{2}(\,j_2^{-}\!-j_1^{-} )&\, \frac{1}{2}(\,j_2^{+}\!-j_1^{+} )&\,( j_1\!-\!j_2)\end{array} \right). & 
\qquad\qquad\qquad\qquad\qquad\;\,{\cal{R}}_3\label{Regge3}
\end{eqnarray}
\begin{eqnarray}
\left(\begin{array}{ccc} j_1 & j_2 & j_3  \\ m_1 & m_2 & m_3 \end{array} \right)
&=&\left(\begin{array}{ccc}\frac{1}{2}(\,j_3^{-}+\,j_2^{-} )&\, \frac{1}{2}(\,j_1^{-}\!+\,j_3^{-} )  &
\, \frac{1}{2}(\,j_2^{-}\!+\,j_1^{-} )\\  [0.25em]
\frac{1}{2}(\,j_3^{-}+\,j_2^{-} )\!-\!j_1^{+} &\, \frac{1}{2}(\,j_1^{-}\!+\,j_3^{-} )\!-\!j_2^{+} &
\, \frac{1}{2}(\,j_2^{-}\!+\,j_1^{-} )\!-\!j_3^{+} \end{array} \right),  
\qquad\quad\,{\cal{R}}_4\label{Regge4}\\ [0.25em]
\phantom{\left(\begin{array}{ccc} j_1 & j_2 & j_3  \\ m_1 & m_2 & m_3 \end{array} \right)}
&=&\left(\begin{array}{ccc}\frac{1}{2}(\,j_3^{+}\!+\,j_2^{+} )& \,\frac{1}{2}(\,j_1^{+}\!+\,j_3^{+} )  &
 \,\frac{1}{2}(\,j_2^{+}\!+\,j_1^{+} )\\  [0.25em]
-\!\frac{1}{2}(\,j_3^{+}\!+\,j_2^{+} )\!+j_1^{-} & -\!\frac{1}{2}(\,j_1^{+}\!+\,j_3^{+} )\!+j_2^{-}&
 -\!\frac{1}{2}(\,j_2^{+}\!+\,j_1^{+} )\!+j_3^{-}
\end{array} \right). 
\quad\;\,{\cal{R}}_5\label{Regge5}\normalsize
\end{eqnarray}
\normalsize

				\subsection{Features of ($3$-$j$) symbols generated by Regge transformations}\label{ReggeFeatures}
We will use various definitions and notations explicited below.\\
$(3$-$j) \stackrel{S_{\boldsymbol{\wr}}}{\longrightarrow}\big\{(3\mbox{-}j)\big\}=\; \mbox{a set denoted by }
{{\tt S}_{\boldsymbol{\wr}}}$
that contains twelve ($3$-$j$). \\
$(3$-$j)\in {{\tt S}_{\boldsymbol{\wr}}}\stackrel{\cal{R}_{\kappa}}{\longrightarrow}(3$-$j)^{\cal{R}_{\kappa} }\in
{{\tt S}_{\boldsymbol{\wr}}^{\cal{R}_{\kappa} }}, \;  \kappa \in [1,5].$\\
Let be $n_{\emptyset}$ the number of empty intersections satisfying to
\begin{equation}
{\tt S}_{\boldsymbol{\wr}}^{\cal{R}_{\kappa} }\cap {\tt S}_{\boldsymbol{\wr}}^{\cal{R}_{\lambda} }\cap 
{\tt S}_{\boldsymbol{\wr}}
=\emptyset , \;  \kappa\neq \lambda \in [1,5].
\end{equation}
{\it A priori} it results that 6 disjoint sets ${{\tt S}_{\boldsymbol{\wr}}}(n_{\emptyset}) $ may be defined for $n_{\emptyset} \in [0,5]$.\\
If a set ${{\tt S}_{\boldsymbol{\wr}}}(n_{\emptyset}) $ is not empty, then it contains $12(n_{\emptyset}+1 )\,(3$-$j)$ symbols.\\
\hspace*{1em}{\tt Filtering operation} $(S_{\boldsymbol{\wr}}\,\mbox{\scriptsize filter})$:\\
${\cal{R}}_{all}$ denotes  the five  Regge transformations. ${\cal{R}}_{all}$ applied to a ($3$-$j$)$_0$
yields a list
\begin{equation}\label{eq:listfive3j}
{\cal{R}}_{all}\big((3\mbox{-}j)_0\big)=\big\{(3\mbox{-}j)^{{\cal{R}}_{1} },
(3\mbox{-}j)^{{\cal{R}}_{2} }, (3\mbox{-}j)^{{\cal{R}}_{3} },
(3\mbox{-}j)^{{\cal{R}}_{4} }, (3\mbox{-}j)^{{\cal{R}}_{5} }\big\}.
\end{equation}
If $(3\mbox{-}j)^{{\cal{R}}_{\lambda} }\in {{\tt S}_{\boldsymbol{\wr}}^{\cal{R}_{\kappa} }}, \lambda \neq \kappa \in [1,5]$
then $(3\mbox{-}j)^{{\cal{R}}_{\lambda} }$ is deleted from the list (\ref{eq:listfive3j}). After this first operation 
there may remain at least {\it one} and at most {\it five} ($3$-$j$) inside the list. 
Among the remaining  ($3$-$j$)'s we continue a similar operation  by checking if a ($3$-$j$) 
$\in {{\tt S}_{\boldsymbol{\wr}_{\,0}}}$,
if it is the case the  ($3$-$j$) is deleted from the remaining list. It may happen that the final list is empty.
The operation described above is denoted by  $(S_{\boldsymbol{\wr}}\,\mbox{\scriptsize filter})$ and we define 
${\cal{R}}_{egge}^{*}$ by\\
\begin{equation}\label{eq:ReggeStarDef}
{\cal{R}}_{egge}^{*}= (S_{\boldsymbol{\wr}}\,\mbox{\scriptsize filter})\circ {\cal{R}}_{all}.
\end{equation}
This allows us to build a partition of any ($3$-$j$) symbols into ${{\tt S}_{\boldsymbol{\wr}}}(n_{\emptyset}) $ sets. 
\begin{equation}
\mbox{ {\tt Closure property}}\; \mbox{under}\; {\cal{R}}_{egge}^{*}\mbox{ is ensured namely} \;
{\cal{R}}_{egge}^{*}\big({{\tt S}_{\boldsymbol{\wr}}}(n_{\emptyset}) \big)
\equiv {{\tt S}_{\boldsymbol{\wr}}}(n_{\emptyset}) . \label{eq:Stability}
\end{equation}
The method is  similar to that followed in our previous 
paper \cite{BrehametPJMPA} about  \{$6$-$j$\}.\\

{\tt Definitions:}\hspace{1em}[{(\tt {circ})} will denote a circular permutation of (1,2,3)]\\
\begin{equation}\label{eq:Nd0}
 {\tt N^{d}_0} =  \mbox{number of zeros of } \big\{( j_i ^{+}\!\! -j_k^{+} )_{i\neq k}\big\}
+  \mbox{number of zeros of } \big\{( j_i ^{-}\!\! -j_k^{-} )_{i\neq k}\big\}.
\end{equation}
\begin{equation}\label{eq:Npm0}
 {\tt N^{\pm}_0} =  \mbox{number of zeros of } \big\{( j_i ^{+}\!\! -j_k^{-} )_{i\neq k}\big\}
+  \mbox{number of zeros of } \big\{( j_i ^{-}\!\! -j_k^{+} )_{i\neq k}\big\}.
\end{equation}
\begin{equation}
 {\tt N^m_0} = \mbox{number of zeros of  }  \big\{( j_i ^{+}\!\! -j_i^{-})\big\}\equiv  \big\{(2m_i)\big\}
\mbox{, with values $0$, $1$ or $3$}.
\end{equation}
Consider $6$ differences between the first row of the Regge array and the second or third.
\begin{equation}\label{eq:ReggeDifferences}
{\tt \delta R_i^k}=(R_1^k\!-\! R_i^k) \quad \mbox{with} \; i \in [2,3], \; k \in [1,3].
\end{equation}
Each quantity is a difference between a $(j^{+}\!- \!j^{-})$ and a  $(j^{-}\!- \!j^{+})$ or a $(j^{+}\!- \!j^{+})$: 
\begin{equation}
{\tt \delta R_2^1} = (j_2^{+}- j_1^{-} ) - (j_1^{-} - j_3^{+}), \quad
{\tt \delta R_3^1} = (j_2^{+}- j_1^{-} ) - (j_1^{+} - j_3^{+})\quad \mbox{and so on}.
\end{equation}
\begin{equation}
{\tt  N^R_{0}} = \mbox{number of zeros of  }[{\tt \delta R_2}] + \mbox{number of zeros of  }[{\tt \delta R_3}] . \qquad
\qquad\quad\;\,
\end{equation}\\ [0.2em]
The partition selectors belong to a set ${\cal{E}}_{\tt Sel}$ ($15$ elements) defined by 
\begin{equation}
{\cal{E}}_{\tt Sel}=\big\{\stackrel{\! \#=3}{( j_i ^{+}\!\! -j_k^{+} )},
\stackrel{\! \#=3}{ ( j_i ^{-}\!\! -j_k^{-} )}, \stackrel{\! \#=3}{( j_i ^{+}\!\! -j_k^{-} )}, 
\stackrel{\! \#=3}{( j_i ^{-}\!\! -j_k^{+} )},  \stackrel{\! \#=3}{( j_i ^{+}\!\! -j_i^{-} )}
\big\}_{i\neq k}\quad i,k \in [1,3] .
\end{equation}
As $[{\stackrel{\! \#=3}{\tt \delta R_2}}], [{\stackrel{\! \#=3}{\tt \delta R_3}}] $ are linear combinations of elements of ${\cal{E}}_{\tt Sel}$, they are not accounted for.\\
{\tt The partitions and selectors found are shown below}. 
\begin{eqnarray}
{\tt S}_{\boldsymbol{\wr}}(0) =\big\{(3\mbox{-}j)\big\}| &
{\tt N^{\pm}_0}\in [3,4]\; {\tt or }\; {\tt N^{\pm}_0}\!=\!6 ,&  \label{eq:S0}\\ [0.5em]
{\tt S}_{\boldsymbol{\wr}}(1)= \big\{(3\mbox{-}j)\big\}| &
{\tt N^{\pm}_0}=2 ,   &
\label{eq:S1}\\ [0.5em]
{\tt S}_{\boldsymbol{\wr}}(2) =\big\{(3\mbox{-}j)\big\}| &
{\tt N^{\pm}_0}=1,  & \label{eq:S2}\\ [0.5em]
{\tt S}_{\boldsymbol{\wr}}(3) =\phantom{ \big\{(3\mbox{-}j)\big\}| } & {\large \emptyset} ,
& \label{eq:S3}\\ [0.5em]
{\tt S}_{\boldsymbol{\wr}}(4) =\big\{(3\mbox{-}j)\big\}|  & 
{\tt N^{\pm}_0}\!=\!0 , {\tt  (N^m_0\!=\!0)} 
 \;\, {\tt and }& \nonumber \\  
&  {\tt  N^{d}_0\!=\!2 , \, N^R_{0}\!=\!0 },\left (
\mathnormal{\big(( j_1^{+} \!=\! j_{2}^{+} )\,{\tt and }\,( j_1 ^{-} \!= \! j_{2}^{-})
\big)}
\,{\tt or}\,{(\tt {circ})} \right)& \nonumber \\
& {\tt or }& \nonumber \\
&{\tt \big(N^{d}_0\!=\!0  ,\, N^R_{0}\!=\!3 )\,{\tt or }\, (N^{d}_0\!=\!4  ,\, N^R_{0}\!=\!0 )\big)} \nonumber\\
& {\tt \oplus }& \nonumber \\
& {\tt N^{\pm}_0}\!=\!0 ,  {\tt  (N^m_0\!=\!1)}
\;\,  {\tt and }\,{\tt (N^d_{0}\!=\!0, N^R_{0}\!=\!4 )}& \nonumber \\ 
& {\tt \oplus }& \nonumber \\
& {\tt N^{\pm}_0}\!=\!0 ,  {\tt  (N^m_0\!=\!3)}\, 
\;\, {\tt and }\,{\tt (N^d_{0}\!=\!0, N^R_{0}\!=\!0 \,{\rm or}\, 2 )},&  \label{eq:S4} \\ [0.5em] 
{\tt S}_{\boldsymbol{\wr}}(5) =\big\{(3\mbox{-}j)\big\}|  & 
{\tt N^{\pm}_0}\!=\!0 ,  {\tt  (N^m_0\!=\!0)} 
\;\, {\tt and }& \nonumber \\ 
& {\tt  N^{d}_0\!=\!2 , \, N^R_{0}\!=\!0 },\left (
\mathnormal{\big(( j_1^{+} \!=\! j_{2}^{+} )\,{\tt and }\,( j_1 ^{-} \neq j_{2}^{-})
\big)}
\,{\tt or}\,{(\tt {circ})} \right)& 
 \nonumber \\ 
& {\tt or }& \nonumber \\
& {\tt \big(N^{d}_0\in [0,1]  ,\, N^R_{0}\in[0,2] \big)\,{\tt or }\, 
 \big( N^{d}_0\!=\!3 , \, N^R_{0}\!=\!0 \big)} \nonumber \\
& {\tt \oplus }& \nonumber \\
& {\tt N^{\pm}_0}\!=\!0 , {\tt  (N^m_0\!=\!1)} 
\;\, {\tt and }\,{\tt \big(N^{d}_0\!=\!0 ,\, N^R_{0}\in[0,2] \big)\,{\tt or }\, 
 \big( N^{d}_0\!=\!1 , \, N^R_{0}\in [0,1] \big)}.
\label{eq:S5} &  
\end{eqnarray}

Instead of $\tt 4$ for \{$6$-$j$\} symbols, we find here $\tt 5$ partitions for ($3$-$j$) symbols.\\
A symbolic sequence illustrate the results where over each subset is indicated its cardinal: 
\begin{equation}
{\tt(3\mbox{-}}j{\tt)}\, {\tt + \, R\!^{^\bigstar}\!\!egge \, symmetry}\longrightarrow \;\,
 \stackrel{\! \#=12}{{\tt {S}}_{\boldsymbol{\wr}}(0) }
\oplus \; \stackrel{\! \#=24}{{\tt {S}}_{\boldsymbol{\wr}}(1) }
\oplus \; \stackrel{\! \#=36}{{\tt {S}}_{\boldsymbol{\wr}}(2) }
\oplus \; \stackrel{\! \#=60}{{\tt {S}}_{\boldsymbol{\wr}}(4) }
\oplus \; \stackrel{\! \#=72}{{\tt {S}}_{\boldsymbol{\wr}}(5) }.
\end{equation}
{\tt As expected the larger symmetry group of order $72, \mbox{{\tt i.e.}}$ $\!\!\stackrel{\! \#=72}{S_{\boldsymbol{\wr}}(5) }$,
is well retrieved, however what remained unknown up to today is the existence of intermediate groups of order $12, 24, 36, 60$
with exclusion of the order 48.}\\ [0.2em]
\hspace*{1em}Achieving this difficult classification requires some comment. Our former program ({\tt symmetryregge})
\cite{BrehametPJMPA} has been modified
into  ({\tt supersymbol3jcount}) where this time  a comparison  of a lot of ${{\tt S}_{\boldsymbol{\wr}}}$ sets is carried out. 
The discoveries of partition selectors are not automatic. Only a thorough examination,  logical or  
intuitive, 
allows one to find them. For lack of a formal logic program able to optimize or reduce possible redundancies, we can not assert that our selectors are the best.
Nevertheless, what is irrefutable is the existence of partitions and selectors. It may be noted also that our results are purely 'computed'  and do not derive
from a group-theoretical analysis (which remains to do).
				\section{Analytic formula for super $(3$-$j)^{S}$ symbols}\label{SuperFormula}
\renewcommand{\theequation}{\ref{SuperFormula}.\arabic{equation}}
\setcounter{equation}{0}
Let us start by updating some definitions used  in a ancient paper  \cite {L.B.Nuov1.2006}.
\begin{equation}   \label{eq:superdelta}
\triangle^{S}(a b c)=\begin{array}{c}
\displaystyle
\left( \frac{[a+b-c]![a-b+c]![-a+b+c]!}{[a+b+c+\frac{1}{2}]!}\right)^{\frac{1}{2}}\end{array}\qquad\mbox{(supertriangle)}.
\end{equation}
Delimiters $\boldsymbol {[}\, \boldsymbol {]}$ around a {\it number}, integer or half-integer, mean 'integer part of {\it number}'. \\
$\bigtriangledown$ stands for $\bigtriangleup ^{-1}$ and $\bigtriangledown^{S}$ for  $\left( {\bigtriangleup^{S}}\right)^{-1}$ . \\
A (so-called) parity independent $(3$-$j)^{S}$  symbol  
\footnote{denoted in \cite{Daumensetal1} by $S3\mbox{-}j$.}
was introduced by Daumens et al. \cite{Daumensetal1} as the product of  
a scalar factor by a standard ($3$-$j$) symbol [its $so(3)$ 'parent']:
\begin{equation}   \label{eq:3jsdefinition}
\left(\!\begin{array}{ccc} j_1\! & j_2\! & j_3 \\
l_1m_1 & l_2m_2 & l_3m_3 \end{array}\! \right)=
\left[ \begin{array}{ccc} j_1 & j_2 & j_3 \\
l_1 & l_2 & l_3 \end{array} \right]
\left(\begin{array}{ccc}l_1 & l_2 & l_3  \\
m_1 & m_2 & m_3 \end{array} \right).
\end{equation}
We have proved   \cite {L.B.Nuov1.2006} that any scalar factor can be written as:
\begin{equation}   \label{eq:ScalarFactor}
\left[ \begin{array}{ccc} j_1 & j_2 & j_3 \\
l_1 & l_2 & l_3 \end{array} \right] = (-1)^{\phi_{\Box}}
\left\{\begin{array}{l} \bigtriangledown(l_1 l_2 l_3)\bigtriangleup^{S}(j_1 j_2 j_3)
\mbox{\hspace{2em}\scriptsize{ $j_1+j_2+j_3$ integer}}\\ 
\bigtriangleup(l_1 l_2 l_3)\bigtriangledown^{S}(j_1 j_2 j_3))
 \mbox{\hspace{1.5em} \scriptsize{$j_1+j_2+j_3$ half-integer}}\end{array} \right. ,
\end{equation}
where the general phase factor $\phi_{\Box}$ can be rewritten as
\begin{equation}   \label{eq:ScalarFactorPhi}
(-1)^{\phi_{\Box}} = (-1)^{2(j_1+j_2+j_3)+8(j_1-l_1)(j_2-l_2)(j_3-l_3)+ 4(l_1(j_3+l_3)+l_2(j_1+l_1)+l_3(j_2+l_2))}.
\end{equation}
We will reuse also our shortened  notation of a $(3$-$j)^{S}$, which drops out  all  $l$'s:
\begin{equation}\label{eq:Short3JSLB}
 \left(\begin{array}{ccc}j_1 & j_2 & j_3  \\ m_1 & m_2 & m_3 \end{array} \right)^{S}\!= 
\left(\!\begin{array}{ccc} j_1\! & j_2\! & j_3 \\ l_1m_1 & l_2m_2 & l_3m_3 \end{array}\! \right).
\end{equation}
It must be realized that the definition of a  super $(3$-$j)^{S}$  implies two triangular constraints,
one for the triangle $(j_1 \,j_2\, j_3)$ with integer or half-integer perimeter, the other for  $(l_1\, l_2\, l_3)$ with integer perimeter only.
For example $| j_1\! -\! j_2| \leq j_3  \leq j_1\!+j_2$ {\sl and} $| l_1\! -\! l_2| \leq l_3  \leq l_1\!+l_2$.\\
It is important for establishing a correct table of $(3$-$j)^{S}$ symbols from (\ref{eq:3jsdefinition}), (\ref{eq:Short3JSLB}) where the 
$l$'s are no more visible and spins $j$ are incremented by step of $\frac{1}{2}$. While forgetting the  condition on the $l$'s, we might have to compute a super symbol like
\scriptsize
$\left(\begin{array}{ccc}7/2 & 2 & 3/2  \\ -1/2 & 1/2 & 0 \end{array} \right)^{S} $, \normalsize
 that has no existence because its parent 
\scriptsize
$\left(\begin{array}{ccc}7/2 & 3/2 & 1  \\ -1/2 & 1/2 & 0 \end{array} \right)$ \normalsize
 is not a valid  $(3$-$j)$ symbol for $so(3)$.
For the calculations now, it seems judicious to gather some square roots together and define a super scalar factor as
\begin{equation}   \label{eq:SuperScalarFactor}
\left[ \begin{array}{ccc} j_1 & j_2 & j_3  \\
m_1 &m_2 &m_3\end{array} \right] ^{S}= 
\bigtriangleup(l_1 l_2 l_3)\left[ \begin{array}{ccc} j_1 & j_2 & j_3 \\
l_1 & l_2 & l_3 \end{array} \right] \qquad\qquad\qquad \mbox{(super scalar factor)}.
\end{equation}
It is of interest because  this super-factor then depends simply of an 
integer positive $\mathrm{I}(j_1 j_2 j_3)$
and of $(j_k\pm m_k)$ for the phase. The result reads
\begin{equation}
\left[ \begin{array}{ccc} j_1 & j_2 & j_3  \\
m_1 &m_2 &m_3\end{array} \right] ^{S}= (-1)^{\phi_{\Box}}\bigtriangleup^{S}(j_1 j_2 j_3)\,\mathrm{\bold I}(j_1 j_2 j_3),
\end{equation}
\begin{equation}\label{eq:Tripair}
\mathrm{\bold I}(j_1 j_2 j_3)= 1\quad \mbox{if $\begin{array}{c}j_1+j_2+j_3\end{array}$= integer},
\end{equation}
\begin{equation}\label{eq:TriImpair}
\mathrm{\bold I}(j_1 j_2 j_3)={ \displaystyle \sum_{k}}(|(-1)^{2(j_k-m_k)}j_k|)\begin{array}{c}+\frac{1}{2}\end{array}
\quad \mbox{if $\begin{array}{c}j_1+j_2+j_3\end{array}$= half-integer}.
\end{equation}
Expression $\sum_k$  is  a trick for representing the four possible positive integer values of $\mathrm{\bold I}$:
\begin{equation}\label{eq:NumberI}\scriptsize
\begin{array}{c}\mathrm{\bold I}_1=(-j_1+j_2+j_3+\frac{1}{2}),\; \mathrm{\bold I}_2=(j_1-j_2+j_3+\frac{1}{2}),\;
\mathrm{\bold I}_3=(j_1+j_2-j_3+\frac{1}{2}),\;\, \mathrm{\bold I}_4=(j_1+j_2+j_3+\frac{1}{2}).
\end{array}\normalsize
\end{equation} 
 Another trick  unifying (\ref{eq:Tripair})-(\ref{eq:TriImpair}) into a single formula is the use of integer parts and factorial.
\begin{equation}\label{eq:NmberI}
\mathrm{\bold I}(j_1 j_2 j_3)=
\frac{\left[{\scriptstyle \left| \sum _{k}(-1)^{2\,{(j_k-m_k)}}\,j_k  \right| +\frac{1}{2}}\right] !}
{\left[{\scriptstyle\left|\sum _{k}(-1)^{2\,{(j_k-m_k)}}\,j_k  \right|}\right] !}.
\end{equation}
An essential remark  concerns the possible doublets $l_k=j_k, l_k=j_k-\frac{1}{2}$. We have
\begin{equation}
(l_k\pm m_k)=\boldsymbol{[}j_k\pm m_k \boldsymbol{]}= \boldsymbol{[}j^{\pm}_k \boldsymbol{]}.
\end{equation}
This gives the means to end all rearrangements and 
adopt a {\tt definition of a   $(3$-$j)^{S}$ symbol fully similar to that of a
$(3$-$j)$ symbol in three equations }like (\ref {eq:Wigner3j})-(\ref{eq:TriangEdmup})-(\ref {eq:v3jformulalb}).
\\ 
\begin{equation}\label{eq:3JSFINALLB}
 \left(\begin{array}{ccc}j_1 & j_2 & j_3  \\ m_1 & m_2 & m_3 \end{array} \right)^{S}=
\triangle^{S}(j_1\, j_2\, j_3)\,
v^{S}\left(\begin{array}{ccc}j_1 & j_2 & j_3  \\
m_1 & m_2 & m_3 \end{array} \right),
\end{equation}
\begin{equation}\label{eq:3JSFINALLB1}
\triangle^{S}(j_1\, j_2\, j_3)=
\begin{array}{c}
\displaystyle
\left( \frac{[j_1+j_2-j_3]![j_1-j_2+j_3]![-j_1+j_2+j_3]!}{[j_1+j_2+j_3+\frac{1}{2}]!}\right)^{\frac{1}{2}}\end{array},
\end{equation}
\begin{eqnarray}\label{eq:3JSFINALLB2}
v^{S}\left(\begin{array}{ccc}j_1 & j_2 & j_3  \\
m_1 & m_2 & m_3 \end{array} \right)= 
(-1)^{\boldsymbol{[}j_1^{+}\boldsymbol{]}-\boldsymbol{[}j_2^{-}\boldsymbol{]}
 +\sum_{k}2j_k+8 \prod_{k}j_k^{\pm}+4(j_1^{\pm}m_2 +j_2^{\pm}m_3 + j_3^{\pm}m_1)}
\scriptsize \begin{array} {c}\left(\prod_{k}\boldsymbol{[}j^{+}_k \boldsymbol{]}! \boldsymbol{[}j^{-}_k\boldsymbol{]}!  \right)
^{\frac{1}{2}}\end{array}\normalsize \qquad\qquad\qquad \nonumber \\
\times 
\scriptsize \frac{\left[{\scriptstyle \left| \sum _{k}(-1)^{2\,j_k^{\pm}}\,j_k  \right| +\frac{1}{2}}\right] !}
{\left[{\scriptstyle\left|\sum _{k}(-1)^{2\,j_k^{\pm}}\,j_k  \right|}\right] !}\normalsize
\begin{array}{c}{\displaystyle \sum_{z}} \frac{(-1)^{z} }
{z!\big(z -(\boldsymbol{[}j^{+}_2 \boldsymbol{]}-\boldsymbol{[}j^{-}_3\boldsymbol{]})\big)!
\big(z-(\boldsymbol{[}j^{-}_1\boldsymbol{]}- \boldsymbol{[}j^{+}_3\boldsymbol{]})\big)!
\big((\boldsymbol{[}j^{+}_1 \boldsymbol{]}+ \boldsymbol{[}j^{+}_2 \boldsymbol{]}- \boldsymbol{[}j^{-}_3 \boldsymbol{]})-z\big)!
\big(\boldsymbol{[}j^{-}_1\boldsymbol{]}-z\big)!\big(\boldsymbol{[}j^{+}_2 \boldsymbol{]}-z\big)!}.
\qquad \end{array}
\end{eqnarray}
For each $(3$-$j)^{S}$,  $so(3) $ doublets can be  retrieved by  using 
$2l_k=\boldsymbol{[}j_k^{+}\boldsymbol{]}+ \boldsymbol{[}j_k^{-}\boldsymbol{]}$.\\
Expressions (\ref{eq:3JSFINALLB})-(\ref{eq:3JSFINALLB2}) allows one to compute a large table of $(3$-$j)^{S}$ that  fits with 
analytic formulas (where one spin equals $\frac{1}{2}$) given in \cite{Daumensetal1}, {\tt after} the correction 
of a misprint\footnote{\cite{Daumensetal1}, p. 2495, Table IV- Analytic values of $S3\mbox{-}j$ symbols, third formula: $\sqrt{\frac{1}{2}}$ to be removed. }.
				\section{Regge symmetry of $(3$-$j)^{S}$ symbols}\label{ReggeSuper}
				\renewcommand{\theequation}{\ref{ReggeSuper}.\arabic{equation}}
				\setcounter{equation}{0}
Exactly as for $\{6$-$j\}^{S}$  \cite{BrehametPJMPA} it is found that $(3$-$j)^{S}$ symbols admit a classification with three intrinsic parities which we will call again $\boldsymbol{\alpha}$, $\boldsymbol{\beta}$, $\boldsymbol{\gamma}$ without confusion with the former ones.\\
\hspace*{1em}Parity $\boldsymbol{\alpha}$:  $(\stackrel{|ev|}{{\tt c}_1}\; \stackrel{|ev|}{{\tt c}_2}\; \stackrel{|ev|}{{\tt c}_3})
^{S}_{\alpha}$.\\ [0.1em]
\hspace*{1em}Parity $\boldsymbol{\beta}$:  $(\stackrel{|ev|}{{\tt c}_1}\; \stackrel{|od|}{{\tt c}_2}\; \stackrel{|od|}{{\tt c}_3})
^{S}_{\beta_1}$ 
 or $(\stackrel{|od|}{{\tt c}_1}\; \stackrel{|ev|}{{\tt c}_2}\; \stackrel{|ev|}{{\tt c}_3})
^{S}_{\beta'_1}$,\\ [0.1em]
\hspace*{5.2em} $(\stackrel{|od|}{{\tt c}_1}\; \stackrel{|ev|}{{\tt c}_2}\; \stackrel{|od|}{{\tt c}_3})
^{S}_{\beta_2}$ 
 or $(\stackrel{|ev|}{{\tt c}_1}\; \stackrel{|od|}{{\tt c}_2}\; \stackrel{|ev|}{{\tt c}_3})
^{S}_{\beta'_2}$,\\ [0.1em]
\hspace*{5.5em}$(\stackrel{|od|}{{\tt c}_1}\; \stackrel{|od|}{{\tt c}_2}\; \stackrel{|ev|}{{\tt c}_3})
^{S}_{\beta_3}$ 
 or $(\stackrel{|ev|}{{\tt c}_1}\; \stackrel{|ev|}{{\tt c}_2}\; \stackrel{|od|}{{\tt c}_3})
^{S}_{\beta'_3}$.\\ [0.1em]
\hspace*{1em}Parity $\boldsymbol{\gamma}$:  $(\stackrel{|od|}{{\tt c}_1}\; \stackrel{|od|}{{\tt c}_2}\; \stackrel{|od|}{{\tt c}_3})
^{S}_{\gamma}$.\\ [0.2em]
Parity $\boldsymbol{\alpha}$ contains only $\begin{array}{c}j_1+j_2+j_3 \end {array}$ integer, $\boldsymbol{\beta}$
can contain $\begin{array}{c}j_1+j_2+j_3 \end {array}$ integer ($\beta_\kappa$)
or half-integer ($\beta'_\kappa $)
and $\boldsymbol{\gamma}$ only $\begin{array}{c}j_1+j_2+j_3 \end {array}$ half-integer. 
Actually this discrepancy is embedded {\it via} the analytic expression of
 $\mathrm{\bold I}(j_1 j_2 j_3)$ given by (\ref{eq:NmberI}), so that
{\tt a best classification of  $(3$-$j)^{S}$ symbols should be expressed in terms of  'column-parity'  and no longer by 
dichotomizing the cases where $\begin{array}{c}\sum_{k} j_k\end {array}$ is integer or half-integer}.\\
According to our defining choice of Regge transformations 
${\cal{R}}_1, {\cal{R}}_2, {\cal{R}}_3, {\cal{R}}_4, {\cal{R}}_5$,
note that \\
$\mathrm{\bold I}_1$ is invariant only under ${\cal{R}}_1$ (parity $\beta_1,\beta'_1 )$,  
$\mathrm{\bold I}_2$ only under ${\cal{R}}_2$ (parity $\beta_2,\beta'_2)$,
$\mathrm{\bold I}_3$ only under ${\cal{R}}_3$ (parity $\beta_3,\beta'_3)$, 
and $\mathrm{\bold I}_4$ under ${\cal{R}}_{all}$ (parity $\gamma )$.
$\mathrm{\bold I}$ numbers were defined by (\ref{eq:NumberI}).\\
\hspace*{1em}A quick reading of the Regge transformations [such as they have been written by (\ref{Regge1})-(\ref{Regge5})]
indicates right away what are the symbols possessing a (super) Regge symmetry.

				\subsection{Features of $(3$-$j)^{S}$  symbols generated by Regge transformations}\label{SuperReggeFeatures}
{\tt Parity }$\boldsymbol{\alpha}, \boldsymbol{\gamma}$:\\
In this case properties like (\ref{eq:S0})-(\ref{eq:S5})  of course are still valid. Thus analogously
\begin{equation}\boxed{
(3\mbox{-}j)^{S}_{\alpha,\gamma} \, {\tt + \, R\!^{^\bigstar}\!\!egge \, symmetry}\longrightarrow \;\,
 \stackrel{\! \#=12}{{\tt {S}}_{\boldsymbol{\wr}}^{S}(0) }
\oplus \; \stackrel{\! \#=24}{{\tt {S}}_{\boldsymbol{\wr}}^{S}(1) }
\oplus \; \stackrel{\! \#=36}{{\tt {S}}_{\boldsymbol{\wr}}^{S}(2) }
\oplus \; \stackrel{\! \#=60}{{\tt {S}}_{\boldsymbol{\wr}}^{S}(4) }
\oplus \; \stackrel{\! \#=72}{{\tt {S}}_{\boldsymbol{\wr}}^{S}(5) }}.
\label{eq:Salphagamma} 
\end{equation}
{\tt Parity} $\boldsymbol{\beta}$\quad  \big(Indices $\kappa \in [1,3]$ of  $\beta_\kappa, \beta'_\kappa$ are no longer significant\big)\\
Only two sets may exist, namely ${\tt {S}}_{\boldsymbol{\wr}}^{S}(0) $ and ${\tt {S}}_{\boldsymbol{\wr}}^{S}(1) $ defined by the selector $\tt N^{\pm}_0$:
\begin{eqnarray}
{\tt S}_{\boldsymbol{\wr}}^{S}(0) =\big\{(3\mbox{-}j)_{\beta}^{S}\big\}| &
{\tt N^{\pm}_0}\in [1,2] , &  \label{eq:SBeta0}\\ [0.5em]
{\tt S}_{\boldsymbol{\wr}}^{S}(1)= \big\{(3\mbox{-}j)_{\beta}^{S}\big\}| &
{\tt N^{\pm}_0}=0 .  &
\label{eq:SBeta1}
\end{eqnarray}
The analog of (\ref{eq:Salphagamma}) then becomes 
\begin{equation}\label{eq:SBetaAnalog} 
\boxed{(3\mbox{-}j)^{S}_{\beta} \, {\tt + \, R\!^{^\bigstar}\!\!egge \, symmetry}\longrightarrow \;\,
 \stackrel{\! \#=12}{{\tt {S}}_{\boldsymbol{\wr}}^{S}(0) }
\oplus \; \stackrel{\! \#=24}{{\tt {S}}_{\boldsymbol{\wr}}^{S}(1) }}.
\end{equation}
\hspace*{1em}{\tt Moreover an unexpected specificity of $\boldsymbol{\beta}$ parity regards the sign of the numerical values of a symbol 
$(3\mbox{-}j)^{S}_{\beta}$ and its transformed by Regge:it can be $\pm$.}\\ [0.2em]
This is explainable by the following proof:
Regge transformations such as described by (\ref{Regge1})-(\ref{Regge5}) and applied formally to a $(3$-$j)^{S}$
leave invariant $\sum_{k}2j_k$. $\forall$ transformation 
$(3\mbox{-}j)^{S}\stackrel{\cal{R}_{\kappa}}{\longrightarrow} (3\mbox{-}j')^{S}$
with $\kappa\in [1,5]$. It can be proved that  only two phases are relevant:
\begin{equation}
(-1)^{\phi^{S}}=
(-1)^{8 \prod_{k}j_k^{\pm}+4(j_1^{\pm}m_2 +j_2^{\pm}m_3 + j_3^{\pm}m_1)}.
\end{equation}
\begin{equation}
(-1)^{\phi'^{S}}=
(-1)^{8 \prod_{k}{j'}_k^{\pm}+4({j'}_1^{\pm}m'_2 +{j'}_2^{\pm}m'_3 + {j'}_3^{\pm}m'_1)}.
\end{equation}
From (\ref{eq:3jsdefinition}) it can be seen that 
\begin{equation}(3\mbox{-}j')^{S}=
(-1)^{\phi^{S}+\phi'^{S}}\times (3\mbox{-}j)^{S}.
\end{equation}
For parities $\boldsymbol{\alpha}$, $\boldsymbol{\gamma}$ we have  $(-1)^{\phi^{S}+\phi'^{S}}=+1$. \\
Consider  a ${{\cal{R}}_{1}}$ transformation, valid for a $(3\mbox{-}j)^{S}_{\beta_{1}}$, we find
a phase $(-1)^{\phi^{S}_{{\cal{R}}_{1}}({\tt c}_1\,{\tt c}_2\,{\tt c}_3)}$ given by
\begin{equation}
(-1)^{\phi^{S}_{{\cal{R}}_{1}}({\tt c}_1\,{\tt c}_2\,{\tt c}_3)}=(-1)^{\phi^{S}_{\beta_1}+\phi'^{S}_{\beta_1}}=
(-1)^{2j_1+4j_1m_1+2j_1^{+}(j_2^{+}- j_3^{+})+((\sum_{k}2j_k)+1)(j_3^{-}-j_2^{-}+1)+2m_2+1}.
\end{equation}
From our definitions of ${\cal{R}}_{1}, {\cal{R}}_{2}, {\cal{R}}_{3}$, it is clear that
\begin{equation}
\phi^{S}_{{\cal{R}}_{2}}({\tt c}_1\,{\tt c}_2\,{\tt c}_3)= \phi^{S}_{{\cal{R}}_{1}}({\tt c}_2\,{\tt c}_1\,{\tt c}_3)
\quad{\mbox{and}}\quad
\phi^{S}_{{\cal{R}}_{3}}({\tt c}_1\,{\tt c}_2\,{\tt c}_3)= \phi^{S}_{{\cal{R}}_{2}}({\tt c}_1\,{\tt c}_3\,{\tt c}_2).
\end{equation}
In shortcut
\begin{equation}
(3\mbox{-}j)^{S}_{\beta_\kappa}\stackrel{{\cal{R}}_{\kappa}}{\longrightarrow} (3\mbox{-}j')^{S}_{\beta_\kappa}
\Longrightarrow(3\mbox{-}j')^{S}_{\beta_\kappa}=(-1)^{\phi^{S}_{{\cal{R}}_{\kappa}}}\times  
(3\mbox{-}j)^{S}_{\beta_\kappa} \quad \mbox{with}\; \kappa \in [1,3].
\end{equation}
{\tt Accordingly, Regge transformations for $\boldsymbol{\beta}$ parity may bring a phase, or not} . \\ 
It depends if $\phi^{S}_{{\cal{R}}_{\kappa}}$ is even or odd.
Tests on computer turn out satisfactory.


				 \section{Analytic prolongation of $(3$-$j)^{S}$ symbols}\label{AnalyticProlong}
				\renewcommand{\theequation}{\ref{AnalyticProlong}.\arabic{equation}}
				\setcounter{equation}{0}

An attempt for extrapolating our table of $(3$-$j)^{S}$ symbols to forbidden cases like $l_3 <| l_1\! -\! l_2|$ or 
 $l_3 >l_1\!+l_2$ produces indefinite values, as expected. It shows 
that only cases of parity $\boldsymbol{\beta}$ are implicated with {\tt flat integer} triangles 
defined by $j_\kappa=j_\lambda+j_\mu \, ,(\kappa,\lambda,\mu)={\tt circ}(1,2,3)$.\\
Let us denote these forbidden cases by {\tt ($3$-$j)^{S\!\boldsymbol{\times}}_{\beta}$} [superscript $\boldsymbol{\times}$ stands for 'forbidden']. 
So to say, they are `orphans' {\it ie} without $so(3)$ parent.  The meaning of scalar factors $\Box$ or integers $\mathrm{\bold I}$ vanishes,
{\it at first sight}. 
For a given $\kappa$, orphan symbols $(3$-$j)^{S\!\boldsymbol{\times}}_\beta$  are precisely 
of the kind $(3$-$j)^{S\!\boldsymbol{\times}}_{\beta_{\kappa}}$.
For example consider $
\scriptsize
\left(\begin{array}{ccc}\stackrel{|od|}{ j_1} & \stackrel{|od|}{j_2} &\stackrel{|ev|}{ j_3 } \\ m_1 & m_2 & m_3 \end{array} 
\right)^{\!\!S\!\boldsymbol{\times}}_{\beta_{3}} 
\normalsize$ where $j_3=j_1+j_2$. 
Clearly $m_1$ can take values varying by a step of $1$: 
$m_1=-j_1+\frac{1}{2}, -j_1+\frac{3}{2},\cdots\!, j_1-\frac{3}{2}, j_1-\frac{1}{2}$. The same holds for $m_2$.
The variation range of $m_3$ is similar, namely: $m_3=-j_3+1, -j_3+2, \cdots\!,j_3-2,j_3-1$. Each increment is $1$.\\
This leads immediately to an analogy with a standard ({\tt flat}) symbol $(3$-$j)$ whose value is derived from a formula given by Edmonds
 \cite[p. 48] {Edmonds}. That reads
\begin{eqnarray}\label{eq:EdmondsFlat}
\left(\begin{array}{ccc} \stackrel{|ev|}{j_1\!-\!\frac{1}{2} }& \stackrel{|ev|}{\,j_2\!-\frac{1}{2}}  
& \stackrel{|ev|}{\,j_1\!+\!j_2\! -\!1 \phantom{\frac{1}{2}}}\\ m_1 & m_2 & m_3 \end{array} \!\!\!\right)=
\begin{array} {c}(-1)^{(j_1\!-\!\frac{1}{2}+m_1)-(j_2\!-\frac{1}{2}-m_2)}\end{array}\quad\quad\quad\nonumber  \\
\times \left[ \! \begin{array}{c} 
\frac{(2j_1-1)!(2j_2-1)!(j_1+j_2-1+m_1+m_2)!(j_1+j_2-1-m_1-m_2)!}
{(2j_1+2j_2-1)!(j_1-\frac{1}{2}+m_1)!(j_1-\frac{1}{2}-m_1)!(j_2-\frac{1}{2}+m_2)!(j_2-\frac{1}{2}-m_2)!}
\end{array}\! \right]^{\frac{1}{2}}.
\end{eqnarray}
From (\ref{eq:ScalarFactor}), after noting that the scalar factor $\left[ \begin{array}{ccc} j_1\!-\!\frac{1}{2} 
& j_2\!-\!\frac{1}{2} & j_3\!-\!1 \\[0.1em]
 j_1\!-\!\frac{1}{2} & j_2\!-\!\frac{1}{2} & j_3\!-\!1 \end{array} \right] 
= \left[ 2j_3\!-\!1\right]^{\frac{1}{2}}$, we  re-write  (\ref{eq:EdmondsFlat}) under a form that highlights our proposal of
analytic prolongation:
\begin{eqnarray}
\left[ \begin{array}{ccc} j_1\!-\!\frac{1}{2} 
& j_2\!-\!\frac{1}{2} & j_3\!-\!1 \\[0.1em]
 j_1\!-\!\frac{1}{2} & j_2\!-\!\frac{1}{2} & j_3\!-\!1 \end{array} \right] 
\left(\begin{array}{ccc}j_1\!-\!\frac{1}{2} &\,j_2\!-\frac{1}{2}  
& \,j_3\!-\!1 \\ m_1 & m_2 & m_3 \end{array} \right)=
\phantom{ \begin{array} {c}(-1)^{(j_1\!-\!\frac{1}{2}+m_1)-(j_2\!-\frac{1}{2}-m_2)}\end{array}} 
\qquad\qquad
\nonumber \\
\begin{array} {c}(-1)^{(j_1\!-\!\frac{1}{2}+m_1)-(j_2\!-\frac{1}{2}-m_2)}\end{array}
\left[ \! \begin{array}{c} 
\frac{(2j_1-1)!(2j_2-1)!(j_3-1+m_3)!(j_3-1-m_3)!}
{(2j_3-2)!(j_1-\frac{1}{2}+m_1)!(j_1-\frac{1}{2}-m_1)!(j_2-\frac{1}{2}+m_2)!(j_2-\frac{1}{2}-m_2)!}
\end{array}\! \right]^{\frac{1}{2}}.
\end{eqnarray}
{\it Analytic prolongation definition}:\\ [0.2em]
In a way fully similar to (\ref{eq:3jsdefinition}), we adopt the following definition, with $j_1,j_2\geq \frac{1}{2}, j_3\geq1$:
\begin{eqnarray}
\left(\begin{array}{ccc}\stackrel{|od|}{ j_1} & \stackrel{|od|}{j_2} &\stackrel{|ev|}{ j_3 } \\ m_1 & m_2 & m_3 \end{array}
 \right)^{\!\!S\!\boldsymbol{\times}}_{\beta_{3} 
\;\,\boldsymbol{|}j_3=j_1+j_2}= 
\left[ \begin{array}{ccc} j_1\!-\!\frac{1}{2} 
& j_2\!-\!\frac{1}{2} & j_3\!-\!1 \\[0.1em]
 j_1\!-\!\frac{1}{2} & j_2\!-\!\frac{1}{2} & j_3\!-\!1 \end{array} \right] 
\left(\begin{array}{ccc}j_1\!-\!\frac{1}{2} &\,j_2\!-\frac{1}{2}  
& \,j_3\!-\!1 \\ m_1 & m_2 & m_3 \end{array} \right)\;\;\, \nonumber \\
=(-1)^{j_1^{+}-j_2^{-}}
 \left[ \! \begin{array}{c} 
\frac{(j_3^{+}-1) ! (j_3^{-}-1)!}
{(2j_3-2)!}\end{array}\! \right]^{\frac{1}{2}}\!\!
\left[ \! \begin{array}{c} 
\frac{(2j_1-1)!(2j_2-1)!}
{\boldsymbol{[}j^{+}_1 \boldsymbol{]}! \boldsymbol{[}j^{-}_1 \boldsymbol{]}!
\boldsymbol{[}j^{+}_2 \boldsymbol{]}! \boldsymbol{[}j^{-}_2 \boldsymbol{]}!}
\end{array}\! \right]^{\frac{1}{2}}.
\end{eqnarray}

Then $(3$-$j)^{S\!\boldsymbol{\times}}
_{\beta_{3} \,\boldsymbol{|}{\tt flat}}$ can be re-integrated in the set of 
regular $(3$-$j)^{S}$ symbols,  according to a single set of equalities
$l_1=j_1\!-\!\frac{1}{2},l_2=j_2\!-\!\frac{1}{2}, l_3=j_3\!-\!1$, by making the following identification 
\begin{eqnarray}
\left(\begin{array}{ccc}\stackrel{|od|}{ j_1} & \stackrel{|od|}{j_2} &\stackrel{|ev|}{ j_3 } \\ m_1 & m_2 & m_3 \end{array}
 \right)^{\!S\!\boldsymbol{\times}}_{\beta_{3} 
\;\,\boldsymbol{|}j_3=j_1+j_2}\simeq
\left(\begin{array}{ccc} \stackrel{|ev|}{j_1\!-\!\frac{1}{2} }& \stackrel{|ev|}{\,j_2\!-\frac{1}{2}}  
& \stackrel{|ev|}{\,j_3\! -\!1 \phantom{\frac{1}{2}} }\\ m_1 & m_2 & m_3 \end{array} \!\!\!\right)
^{\!\!S}_{\boldsymbol{\alpha} 
\;\,\boldsymbol{|}j_3=j_1+j_2}.
\end{eqnarray}
More generally
\begin{eqnarray}\label{eq:ExtendBetaForbid}
\left(\begin{array}{ccc} j_1 & j_2 & j_3  \\ m_1 & m_2 & m_3 \end{array} \right)^{\!S\!\boldsymbol{\times}}
_{\beta_{\kappa} \,\boldsymbol{|}{\tt flat}}\simeq
\left(\begin{array}{ccc}j_\lambda\!-\!\frac{1}{2} &\,j_\mu\!-\frac{1}{2}  
& \,j_\kappa\!-\!1 \\ m_\lambda & m_\mu & m_\kappa \end{array} \right)
^{\!\!S}_{ \boldsymbol{\alpha}\;\, \scriptsize
\begin{array}{|l}\, j_\kappa=j_\lambda+j_\mu\, j_\lambda,j_\mu\geq \frac{1}{2} \\ \, (\kappa,\lambda,\mu)={\tt circ}(1,2,3) \end{array} }
,
\end{eqnarray}
with the following numerical value
\begin{eqnarray}\label{eq:3JSFINALLB2PHANTOMGeneral2}
\left(\begin{array}{ccc} j_1 & j_2 & j_3  \\ m_1 & m_2 & m_3 \end{array} \right)^{\!S\!\boldsymbol{\times}}
_{\beta_{\kappa} \,\boldsymbol{|}{\tt flat}}
= 
(-1)^{j_\lambda^{+}-j_\mu^{-}}
  \left[ \! \begin{array}{c} 
\frac{(j_\kappa^{+}-1)! (j_\kappa^{-}-1)!}
{(2j_\kappa-2)!}\end{array}\! \right]^{\frac{1}{2}}\!
\left[ \!\begin{array}{c} 
\frac{(2j_\lambda-1)!(2j_\mu-1)!}
{\boldsymbol{[}j^{+}_\lambda \boldsymbol{]}! \boldsymbol{[}j^{-}_\lambda\boldsymbol{]}!
\boldsymbol{[}j^{+}_\mu \boldsymbol{]}! \boldsymbol{[}j^{-}_\mu\boldsymbol{]}!}
\end{array}\!\right]^{\frac{1}{2}}
.\end{eqnarray}

{\tt Regge transformations and notation for flat triangles}:\\
Since a symbol
$(3\mbox{-}j)^{S\!\boldsymbol{\times}}_{\beta_{\kappa}\;\,\boldsymbol{|}{\tt flat }}$  is actually of the kind $\boldsymbol{\alpha}$
then ${\cal{R}}_{all}\left((3\mbox{-}j)^{S\!\boldsymbol{\times}}_{\beta_{\kappa}\;\,\boldsymbol{|}{\tt flat }}\right)$ have their five identical numerical values, phase included. In order to ensure {\tt the closure property} (\ref {eq:Stability}),
we need an additional filtering operation $(S_{\boldsymbol{\wr}}\,\mbox{\underline{\scriptsize filter}})$, where the bar which underlines 
means that only flat triangles $(j_1 j_2 j_3)$ are retained. Extension of this underlining will be used elsewhere with an obvious signification.
Analogously to (\ref {eq:ReggeStarDef}) we may define a ${\cal{\underline{R}}}_{\,egge}^{*}$ as 
\begin{equation}
{\cal{\underline{R}}}_{\,egge}^{*}= (S_{\boldsymbol{\wr}}\,\mbox{\underline{\scriptsize filter}})\circ
(S_{\boldsymbol{\wr}}\,\mbox{\scriptsize filter})\circ {\cal{R}}_{all}.
\end{equation}
Clearly the number of disjoint sets ${{\tt \underline{S}}\,^{S}_{\wr}}(n_{\emptyset}) $ will be reduced. A bit like for a true parity
$\boldsymbol{\beta}$ ({\it ie} valid),
the remaining selection comes from only one ${{\cal{R}}_{1}}$, or ${{\cal{R}}_{2}}$ or 
${{\cal{R}}_{3}}$. Accordingly both possible values of $n_{\emptyset} $ belong to the range $[0,1]$. We can present the results as follows:
\begin{equation}\label{eq:ReggeForbidden}
\boxed{
(3\mbox{-}j)^{S\!\boldsymbol{\times}}_{\underline{\beta}}\simeq
(3\mbox{-}j)^{S}_{\underline{\alpha}}
\;\; {\tt \boldsymbol{+} \;\; \underline{R}\!^{^\bigstar}\!\!egge \, symmetry}\longrightarrow \;\,
 \stackrel{\! \#=12}{{\tt {\underline{S}}}\,_{\wr}^{S}(0) }
\oplus \; \stackrel{\! \#=24}{{\tt {\underline{S}}}\,_{\wr}^{S}(1) }}.
\end{equation}
The relevant selectors here  and their notations are slightly different from (\ref{eq:Nd0})-(\ref{eq:Npm0}).
\begin{equation}
 {\tt \underline{N}^{\,d}_{\,0}}
 =  \mbox{number of zeros of } \big\{( \boldsymbol{\jmath}_i ^{+}\!\! -\boldsymbol{\jmath}_k^{+} )_{i\neq k}\big\}
+  \mbox{number of zeros of } \big\{( \boldsymbol{\jmath}_i ^{-}\!\! -\boldsymbol{\jmath}_k^{-} )_{i\neq k}\big\},
\end{equation}
\begin{equation}
 {\tt \underline{N}^{\,\pm}_{\,\,0}} =  \mbox{number of zeros of } \big\{( \boldsymbol{\jmath}_i ^{+}\!\! -\boldsymbol{\jmath}_k^{-} )_{i\neq k}\big\}
+  \mbox{number of zeros of } \big\{( \boldsymbol{\jmath}_i ^{-}\!\! -\boldsymbol{\jmath}_k^{+} )_{i\neq k}\big\},
\end{equation}
where the spins $\boldsymbol{\jmath}$ are defined from (\ref{eq:ExtendBetaForbid}) by
\begin{eqnarray}
\boldsymbol{\jmath}_\lambda=\begin{array}{c}j_{\lambda}\!-\!\frac{1}{2}\end{array},
& \boldsymbol{\jmath}_\mu=\begin{array}{c}j_{\mu}\!-\!\frac{1}{2}\end{array},
& \boldsymbol{\jmath}_\kappa=\begin{array}{c}j_{\kappa}\!-\!1\end{array}.
\end{eqnarray}
Below are listed  the selectors and their values such as we found them:
\begin{eqnarray}
{\tt \underline{S}}\,_{\wr}^{S}(0) =\big\{(3\mbox{-}j)_{\underline{\beta_\kappa}}^{S\!\boldsymbol{\times}}\big\}| &
 {\tt \underline{N}^{\,\pm}_{\,\,0}}=1,{\tt \underline{N}^{\,d}_{\,0}}\in [0,2] ,
 \big(( \boldsymbol{\jmath}_\lambda ^{+}\!\! =\boldsymbol{\jmath}_\mu^{-} )\,{\tt or}\,
( \boldsymbol{\jmath}_\lambda ^{-}\!\! =\boldsymbol{\jmath}_\mu^{+} )\big)
& \nonumber \\ 	
& {\tt or }& \nonumber \\
&  [{\tt \underline{N}^{\,\pm}_{\,\,0}}=2],({\tt \underline{N}^{\,d}_{\,0}}=0 \,{\tt or}\, 2 ),
( \boldsymbol{\jmath}_\lambda ^{+}\!\! =\boldsymbol{\jmath}_\mu^{-} )\,{\tt and}\,
( \boldsymbol{\jmath}_\lambda ^{-}\!\! =\boldsymbol{\jmath}_\mu^{+} )
& \nonumber \\   
& {\tt or}\; ({\tt \underline{N}^{\,d}_{\,0}}=1 \,{\tt or}\, 3 ),
\big(( \boldsymbol{\jmath}_\lambda ^{+}\!\! =\boldsymbol{\jmath}_\mu^{-}\!\! =\boldsymbol{\jmath}_\kappa^{+} )
\,{\tt or}\,( \boldsymbol{\jmath}_\lambda ^{-}\!\! =\boldsymbol{\jmath}_\mu^{+}\!\! =\boldsymbol{\jmath}_\kappa^{-} )\big)
& \nonumber \\   
& {\tt or }& \nonumber \\
& {\tt \underline{N}^{\,\pm}_{\,\,0}}\in [3,4]\; {\tt or }\; {\tt \underline{N}^{\,\pm}_{\,\,0}}\!=\!6
 , &  \label{eq:SBetaX0}\\ [0.5em]
{\tt \underline{S}}\,_{\wr}^{S}(1)= \big\{(3\mbox{-}j)_{\underline{\beta_\kappa}}^{S\!\boldsymbol{\times}}\big\}| &
  {\tt \underline{N}^{\,\pm}_{\,\,0}}=0 \nonumber \\
& {\tt or }& \nonumber \\
&{\tt \underline{N}^{\,\pm}_{\,\,0}}=1,{\tt \underline{N}^{\,d}_{\,0}}\in [0,2] \, {\tt and} & \nonumber \\
& \big(( \boldsymbol{\jmath}_\lambda ^{+}\!\! =\boldsymbol{\jmath}_\kappa^{-} )\,{\tt or}\,
( \boldsymbol{\jmath}_\lambda ^{-}\!\! =\boldsymbol{\jmath}_\kappa^{+} )\, {\tt or} \,
( \boldsymbol{\jmath}_\mu ^{+}\!\! =\boldsymbol{\jmath}_\kappa^{-} )\,{\tt or}\,
( \boldsymbol{\jmath}_\mu ^{-}\!\! =\boldsymbol{\jmath}_\kappa^{+} )\big)
& \nonumber \\ 	
& {\tt or }& \nonumber \\
& [{\tt \underline{N}^{\,\pm}_{\,\,0}}=2], & \nonumber \\
 &{\tt \underline{N}^{\,d}_{\,0}}=0 , \big(( \boldsymbol{\jmath}_\lambda ^{+}\!\! =\boldsymbol{\jmath}_\kappa^{-} )\,{\tt and}\,
( \boldsymbol{\jmath}_\lambda ^{-}\!\! =\boldsymbol{\jmath}_\kappa^{+} )\big)\, {\tt or}\,
 \big(( \boldsymbol{\jmath}_\mu ^{+}\!\! =\boldsymbol{\jmath}_\kappa^{-} )\,{\tt and}\,
( \boldsymbol{\jmath}_\mu ^{-}\!\! =\boldsymbol{\jmath}_\kappa^{+} )\big)\, & \nonumber  \\
& {\tt or}\;{\tt \underline{N}^{\,d}_{\,0}}=1,
\,\big(( \boldsymbol{\jmath}_\lambda ^{+}\!\! =\boldsymbol{\jmath}_\mu^{+}\!\! =\boldsymbol{\jmath}_\kappa^{-} )
\,{\tt or}\, ( \boldsymbol{\jmath}_\lambda ^{-}\!\! =\boldsymbol{\jmath}_\mu^{-}\!\! =\boldsymbol{\jmath}_\kappa^{+} ) \big)
\phantom{{\tt \underline{N}^{\,d}_{\,0}}=1,}\qquad\qquad\qquad\;\;
 &\nonumber \\
&{\tt or}\;{\tt \underline{N}^{\,d}_{\,0}}=2 , \big(( \boldsymbol{\jmath}_\lambda ^{+}\!\! =\boldsymbol{\jmath}_\kappa^{-} )
\,{\tt and}\,
( \boldsymbol{\jmath}_\lambda ^{-}\!\! =\boldsymbol{\jmath}_\kappa^{+} )\big)\, {\tt or}\,
 \big(( \boldsymbol{\jmath}_\mu ^{+}\!\! =\boldsymbol{\jmath}_\kappa^{-} )\,{\tt and}\,
( \boldsymbol{\jmath}_\mu ^{-}\!\! =\boldsymbol{\jmath}_\kappa^{+} )\big)\;\,\quad\nonumber \\
& \phantom{{\tt \underline{N}^{\,d}_{\,0}}=2,}\,{\tt or} 
\,\big(( \boldsymbol{\jmath}_\lambda ^{+}\!\! =\boldsymbol{\jmath}_\mu^{+}\!\! =\boldsymbol{\jmath}_\kappa^{-} )
\,{\tt or}\, ( \boldsymbol{\jmath}_\lambda ^{-}\!\! =\boldsymbol{\jmath}_\mu^{-}\!\! =\boldsymbol{\jmath}_\kappa^{+} ) \big).
\qquad\qquad\qquad\;\, 
 &
   \label{eq:SBetaX1}
\end{eqnarray}
Note that $( \boldsymbol{\jmath}_\lambda ^{\pm}\!\! =\boldsymbol{\jmath}_\mu^{\mp})\equiv
( j_\lambda ^{\pm}\!\! =j_\mu^{\mp})$ and $( \boldsymbol{\jmath}_\lambda ^{\pm}\!\! =\boldsymbol{\jmath}_\mu^{\pm})\equiv
( j_\lambda ^{\pm}\!\! =j_\mu^{\pm})$.
Again, it is awkward to have so many defining equations of selectors for a few flat triangles. 
All could certainly be simplified and presented otherwise  by optimizing the selectors that we have adopted throughout the research. 
This is another matter for reflection, not discussed in this study.\\
\hspace*{1em}The advantage of the proposed analytical extension allows one to compute a complete table of $(3$-$j)^{S}$ symbols
where the spins can vary by step of $\frac{1}{2}$ by considering only the triangular constraint on the triangle $(j_1\,j_2\,j_3)$.

				\section{Conclusion}\label{conclusion}
				\renewcommand{\theequation}{\ref{conclusion}.\arabic{equation}}
				\setcounter{equation}{0}

 As known the set of $\sigma$-orbits can provide a partition of a symmetric group $S_k$, however
the present situation  is different since a partition of any $(3$-$j)$ or $(3$-$j)^{S}$  symbol is built from linear transformations (Regge).
Although the $(3$-$j)$ symbols are  objects simpler than the $\{6$-$j\}$ symbols \cite{Granovskii}, for now this feature is far from evident when considering the disparities between the (Regge) partitions found here and that of $\{6$-$j\} / \{6$-$j\}^{S}$ analyzed in \cite{BrehametPJMPA}.
Our partitions and selectors being properly identified, the ideal would be to derive those of  $(3$-$j) / (3$-$j)^{S}$ from those 
of $\{6$-$j\} / \{6$-$j\}^{S}$ since it is recognized that a $(3$-$j)$ may be viewed as an asymptotic limit of a $\{6$-$j\}$. 
``{\it In  this limit the $6j$ Regge symmetry becomes the $3j$ Regge symmetry}'' \cite[p. 118] {Boalch}.\\
\hspace*{1em}Actually the results obtained in this paper are far to close the studies of Regge symmetries sometimes qualified of 'mysterious' or 'surprising', 
even after the most recent studies \cite{Granovskii,Boalch}. Another approach oriented to the point of view of partitions might be fruitful. 


\end{flushleft}
\end{document}